# Y(Ni, Mn)O$_3$ epitaxial thin films prepared by pulsed laser deposition

Yanwei Ma[1], M. Guilloux-Viry[1], O. Pena[1], C. Moure[2]

[1] Chimie du Solide et Inorganique Moleculaire (LCSIM), UMR-CNRS 6511, Universite de Rennes 1, Institut de Chimie, 35042 Rennes, France

[2] Instituto de Ceramica y Vidrio, CSIC, 28049 Madrid, Spain

**Abstract**

High–quality epitaxial YNi$_x$Mn$_{1-x}$O$_3$ thin films have been successfully grown on SrTiO$_3$ (100) (STO) by pulsed laser deposition. X-ray diffraction studies showed that the films deposited on STO are fully c-axis oriented and exhibit in-plane alignment. The magnetic transition temperatures ($T_c$) of the films (both x=0.33 and 0.5) are equivalent to the values of the corresponding bulk samples. However, when x=0.5, the films show magnetic properties quite different from those of bulk samples. This difference may be caused by the structure distortion in these films.

**Introduction**

Doped perovskite manganites $RE_{1-x}A_xMnO_3$ (RE=rare-earth metal, A=divalent element) have been extensively studied over the last decade. This system is found to exhibit fascinating properties [1,2], such as metal-insulator transition (MIT), ferromagnetic (FM) -paramagnetic (PM) phase transition, colossal magnetoresistance (CMR), charge and orbital ordering, etc., which can be very useful for the development of new magnetic and magnetoresistive devices. Previously, most of these works report on the substitutions of the rare-earth ion (A-site of the $ABO_3$ perovskite) by an alkaline-earth element. In such a case, each time a divalent alkaline earth element $M^{2+}$ replaces one trivalent cation, one atom of $Mn^{3+}$ transforms into $Mn^{4+}$, i.e. $RE_{1-x}^{3+}A_x^{2+}Mn_{1-x}^{3+}Mn_x^{4+}O_3^{2-}$, and the magnetic and transport properties of the manganites change. The mixed valency of Mn ions leads to strong ferromagnetic (FM) interaction arising from the $Mn^{3+}$-O-$Mn^{4+}$ bonds. In general, the transport and magnetic properties of the manganites were explained by the double-exchange (DE) mechanism [3] combining with the local Jahn-Teller distortions of $Mn^{3+}$ ions [4].

On the other hand, there are only a few reports on doping the B-site of the $ABO_3$ perovskite, in which manganese can be partially substituted by divalent transition elements (e.g., $Cu^{2+}$, $Co^{2+}$, $Ni^{2+}$…). When a divalent ion $M^{2+}$ substitutes one manganese atom in $RE(M,Mn)O_3$, a transformation mechanism $Mn^{3+} \rightarrow Mn^{4+}$ takes place, leading to the $RE^{3+}M_x^{2+}Mn_{1-2x}^{3+}Mn_x^{4+}O_3^{2-}$ formulation. This time however, two Mn atoms are involved in the transformation, one being replaced by the substituent, the second one transforming into $Mn^{4+}$. As a consequence, it is expected that DE interactions and electrical-conduction mechanisms should be optimized at the $x = x_{crit} = 1/3$ substituent concentration, for which an equal number of $Mn^{3+}$ and $Mn^{4+}$ ions coexist in the material.

In previous works concerning this type of substitution in the solid solution Y(Ni,Mn)$O_3$, we have reported the effect of Ni substitution for Mn on the structural, electrical and magnetic properties of the Y-based manganite system YNi$_x$Mn$_{1-x}$O$_3$ (YNMO), and found that the magnetic regime depends on the relative concentration of the substituent, changing from antiferromagnetism (x < 0.33) to ferromagnetism (x > 0.33) [5-6]. Moreover, the electrical and magnetic properties of manganite thin films are often very different than those of the materials produced by bulk ceramic techniques or single crystals with the same nominal composition.

In order to integrate these functional materials into technological devices, it is of first importance, in a first step, to study their behavior in the form of thin films. In this respect, we have undertaken the elaboration of thin films of YNMO, for some specific x(Ni) compositions (x = 0.33 and 0.50). Here we report the experimental results of the structural and magnetic measurements on the YNMO thin films, such as XRD, the dependence of magnetization on temperature as well as on applied fields.

**Experimental**

The pulsed laser deposition (PLD) apparatus used is described elsewhere [7]. Sintered polycrystalline YNi$_x$Mn$_{1-x}$O$_3$ targets were ablated towards to single-crystal substrates of (100) SrTiO$_3$ (STO) by a KrF excimer laser (Tuilasea wavelength: 248 nm) with a fluence of 2 J/cm$^2$ at 2 Hz. SrTiO$_3$ is a standard substrate used for the deposition of functional perovskite-like compounds because of its perovskite structure and its commercial availability. In order to optimize the growth conditions, we have studied the dependence of the transition temperature (Tc) on the substrate temperature (Ts) and on the oxygen pressure during deposition. We found that at the substrate temperature of 740°C and in the oxygen pressure range of 0.25-0.6 mbar, the Tc of the film is close to that of the target material. After deposition was stopped, the oxygen pressure was increased to a static value of 200 Torr, and the films were cooled to room temperature at ~35 °C/min.

The film structures were examined by x-ray diffraction (XRD) using a high-resolution four-circle texture diffractometer (Brüka AXS, D8 Discover). The magnetization of the thin films was measured using a SQUID magnetometer (MPMS-XL5, Quantum Design). Thin film samples were held in a plastic straw. Measured data contained contributions from the film and the STO substrate which is diamagnetic. All magnetization measurements were performed in magnetic fields applied parallel to the film surface. For each M-T curve, the sample was cooled down to 6 K in zero magnetic field, and the magnetization was measured by warming the sample in an applied field (ZFC). Then, the magnetization was obtained while the sample was being cooled in the same magnetic field (FC).

**Results and discussion**

The structure of the bulk YNMO was found to be orthorhombic (*Pbnm*). For x=0.5, the lattice constants are $a/\sqrt{2}= 3.692$ Å, $b/\sqrt{2}= 4.007$ Å, $c/2=3.706$ Å. As expected for epitaxial films, the in-plane lattice parameters *a* and *b* match the substrate of SrTiO$_3$ (cubic, $a=3.905$ Å) provided that the film crystal lattice is rotated by 45° with respect to the substrate. Therefore, the YNMO/STO films have the film-substrate lattice mismatch ~ 5.4% along *a* direction, ~-2.6% along *b* direction, with their in-plane lattice parameters expanded and compressed, respectively.

Figure 1 shows the typical x-ray θ/2θ scan recorded for a film of YNi$_{0.33}$Mn$_{0.67}$O$_3$ and YNi$_{0.5}$Mn$_{0.5}$O$_3$, respectively. As can be seen, only the reflections from planes perpendicular to the substrate are observed, i.e. $h=k=0$, $l\neq0$, indicating that the films were grown preferentially with the *c* axes normal to the film plane. No impurity phases were found. In order to investigate the crystal quality of these YNMO films, the rocking curves of the (004) peaks were explored by θ-scan. The full width at half maximum (FWHM) of both films is around 1°. Furthermore, the four peaks at 90° intervals in the φ-scan of the x=0.33 film (Fig. 2) make evident the existence of an in-plane order of the film. The same was found to be the case for x=0.5. These observations confirmed the high crystalline quality and good epitaxy of the YNMO thin films.

Previous magnetic measurements on bulk samples of $YNi_xMn_{1-x}O_3$ showed the existence of a critical concentration $x_{crit}$ ($x_{crit} = 1/3$), which optimizes both the double-exchange magnetic interactions and the controlled-valence electrical-conduction mechanism. At high Ni concentration (x=0.50), the system is ferromagnetic with a critical temperature of ~ 80 K. The hysteresis cycles present a small hysteresis of about 200 Oe (Hc) and a magnetization of 2.1 $\mu_B$ per Mn atom [6].

In order to compare the magnetic behavior of our thin films with that observed in bulk samples, field-cooled (FC) and zero-field-cooled (ZFC) magnetization curves M(T) were measured at a low field of 100 Oe, as shown in Fig.3. For x=0.5, the Tc value is ~ 80 K, very close to the reported value of the corresponding bulk samples [6]. The relatively broad magnetic transition (Tc) of this film may be attributed to possible magnetic inhomogeneity which, in turn, may be due to nonuniform distribution of grains with different size and/or oxygen content in the film. At low temperatures, the ZFC magnetization M shows more or less constant, then increases rapidly above ~60 K and shows a peak at a temperature of ~72 K (also defined as the freezing temperature of clusters [8]), which shifts to lower temperatures for higher applied fields. On the contrary, $M_{FC}$ continuously increases below $T_{irr}\sim76$ K resulting in large difference between FC and ZFC magnetizations. Clearly, this magnetization behavior is quite different from that of bulk samples with same composition, which showed an almost constant magnetization below Tc ~80 K.

The large irreversibility of this sample between the ZFC and FC is distinct, suggesting the existence of magnetically ordered small regions separated by a matrix of disordered spins, which is usually treated as a ''cluster'' glass state [8]. It was claimed that a cluster glass phase has no simple long-range FM order. When x=0.33, a clear cusp appears in the ZFC magnetization (Fig.3b), indicating that the cluster glass (x=0.5) is mostly changed into the spin glass state (x=0.33).

As seen in Fig.3b, the x=0.33 film shows a spin canting-like transition around 36 K, getting broader and moving downwards in temperature with increasing field (not shown), very similar to that of the bulk samples of the same composition. This behavior is believed to be characteristic of canted-antiferromagnetism, in which a ferromagnetic component coexists with predominant antiferromagnetic interactions.

Figure 4 presents the field dependence of magnetization for x=0.33 and 0.5 at 5 K. The magnetization of the x=0.5 film tends to saturate at rather low fields, of the order of 20 kOe, with a usual hysteresis loop below 10 kOe. However, this hysteresis is much bigger than that of the bulk sample, indicating large coercive fields (Hc=1500 Oe and 200 Oe for film and bulk samples, respectively). The large value of the coercive field for x=0.5 further confirm the presence of magnetic clusters, that is, clusters randomly freeze in locally canted states at low temperatures, for which the random anisotropy energy will act as pinning potential [9].

On the contrary, for x=0.33, the magnetization initially increases rapidly but does not saturate even at the higher applied field. The data suggest the existence of both a ferromagnetic and an antiferromagnetic components. By increasing the applied field, the ferromagnetic part tends to saturate, whereas the antiferromagnetic part increases linearly, resulting in lack of saturation of magnetization.

From the data mentioned above, the magnetic properties of x=0.33 thin films are bulk-like, and compare well with data reported in ref. [6]. However, the x=0.5 films exhibited much different magnetic properties compared with the corresponding bulks. The difference in magnetic behavior between film and bulk samples may be caused by the structure distortion of the thin film due to lattice mismatch between the YNMO film and the substrate STO. The appearance of cluster glass, indicative of a magnetic disorder, is the evidence for weakening of the DE interaction caused by the structural distortion.

**Conclusions**

The structural and magnetic properties of $YNi_xMn_{1-x}O_3$ films deposited on (100) $SrTiO_3$ were investigated for the first time. X-ray diffraction studies show that the films are single phase and epitaxially grown along the (00*l*) orientation. The magnetic transition temperatures ($T_c$) of the films are equivalent to the values of the corresponding bulk samples. Magnetic measurements suggest a spin-glass characteristic in the x=0.33, while a cluster glasslike behavior is observed for x=0.5, which is quite different from that of the bulk sample. This difference may be associated with a different magnetic structure in both films due to the presence of locally canted states in the case of the x=0.5 films.

**Acknowledgments**

One author (Y. M.) thanks Dr. M. Bahout for helpful discussions and T. Guizouarn for technical assistance during measurements. Financial support from Region Bretagne is greatly acknowledged.

# References

[1] R. von Helmolt, J. Wecker, B. Holzapfel, L. Schultz, and K. Samwer, Phys. Rev. Lett.71, 233 (1993).

[2] S. Jin, H. M. O'Bryan, T. H. Tiefel, M. McCormack, and W. W. Rhodes, Appl. Phys. Lett. 66, 382 (1995).

[3] C. Zener, Phys. Rev. 82, 403 (1951).

[4] A. J. Millis, P. B. Littlewood, and B. I. Shraiman, Phys. Rev. Lett. 74, 5144 (1995).

[5] C. Moure, D. Gutierrez, O. Pena, P. Duran, J. Solid State Chem.163 (2002) 377.

[6] O. Pena, M. Bahout, D. Gutierrez, J.F. Fernandez, P. Duran, C. Moure, J. Phys. Chem. Solids 61 (2000) 2019.

[7] J. R. Duclère, M. Guilloux-Viry, A. Perrin, E. Cattan, C. Soyer, and D. Rèmiens, Appl. Phys. Lett. 81, 2067 (2002).

[8] M. A. Señarís-Rodríguez and J. B. Goodenough, J. Solid State Chem.118 (1995) 323.

[9] P. Nozar, V. Sechovsky and V. Kambersky, J. Magn. Magn. Mater. 69, 71 (1987) .

**Captions**

Figure 1 θ-2θ XRD patterns of YNMO films grown on STO substrates for x=0.33 and 0.5, respectively.

Figure 2 X-ray φ scan of the (111) reflection peak of the YNMO film (2θ=25.92°, ψ=62.61°). The fourfold symmetry reveals that the film deposited on STO is good in-plane order. Note: two small peaks of 2θ=38.286° and 44.423° were contributed from the sample holder.

Figure 3 Magnetization as a function of temperature measured at a low field of 100 Oe for x=0.33 and 0.5 films.

Figure 4 Hysteresis loops at 5 K up to 30 kOe of thin films (a) x=0.33; (b) x=0.5.

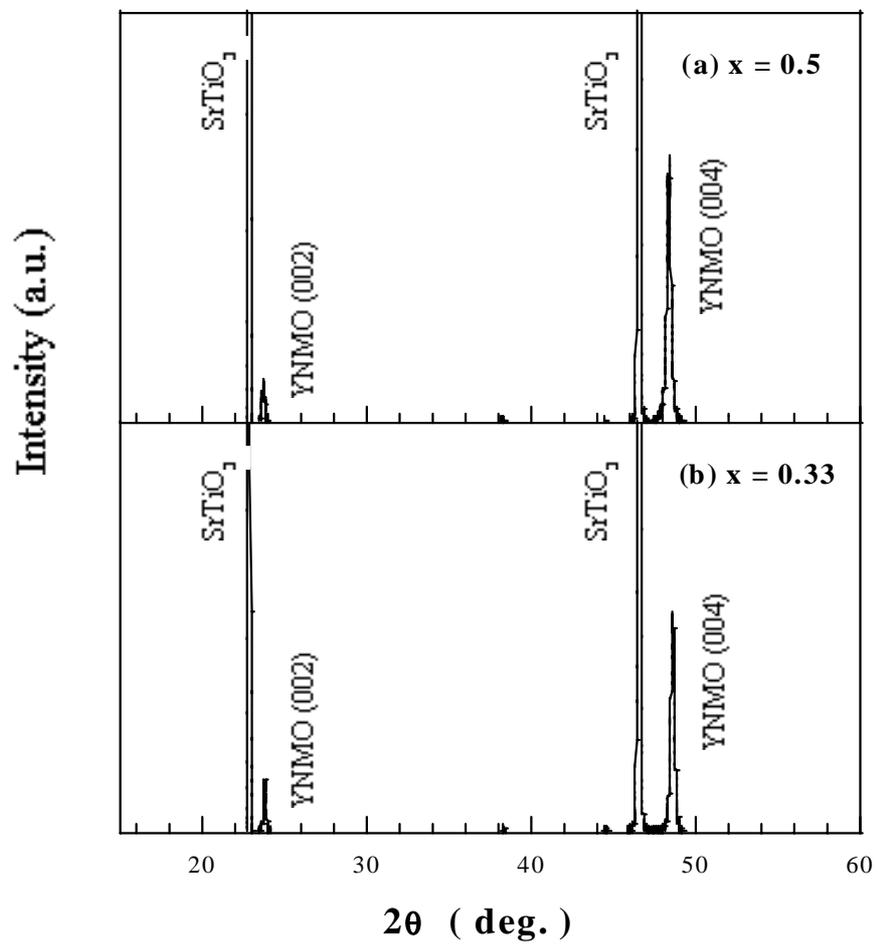

Fig. 1   Ma et al.

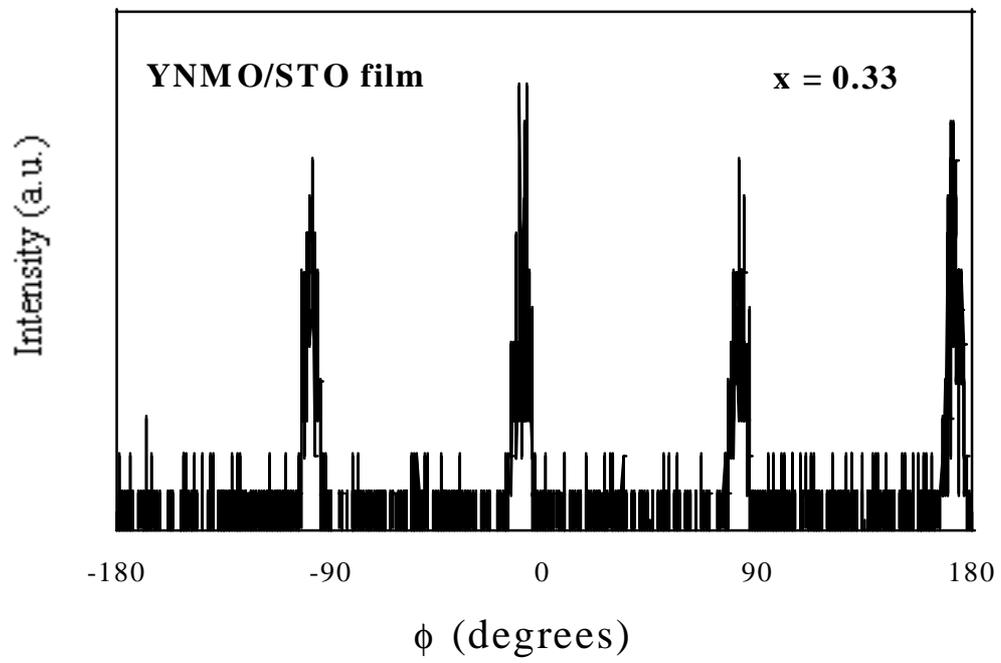

Fig. 2  Ma et al.

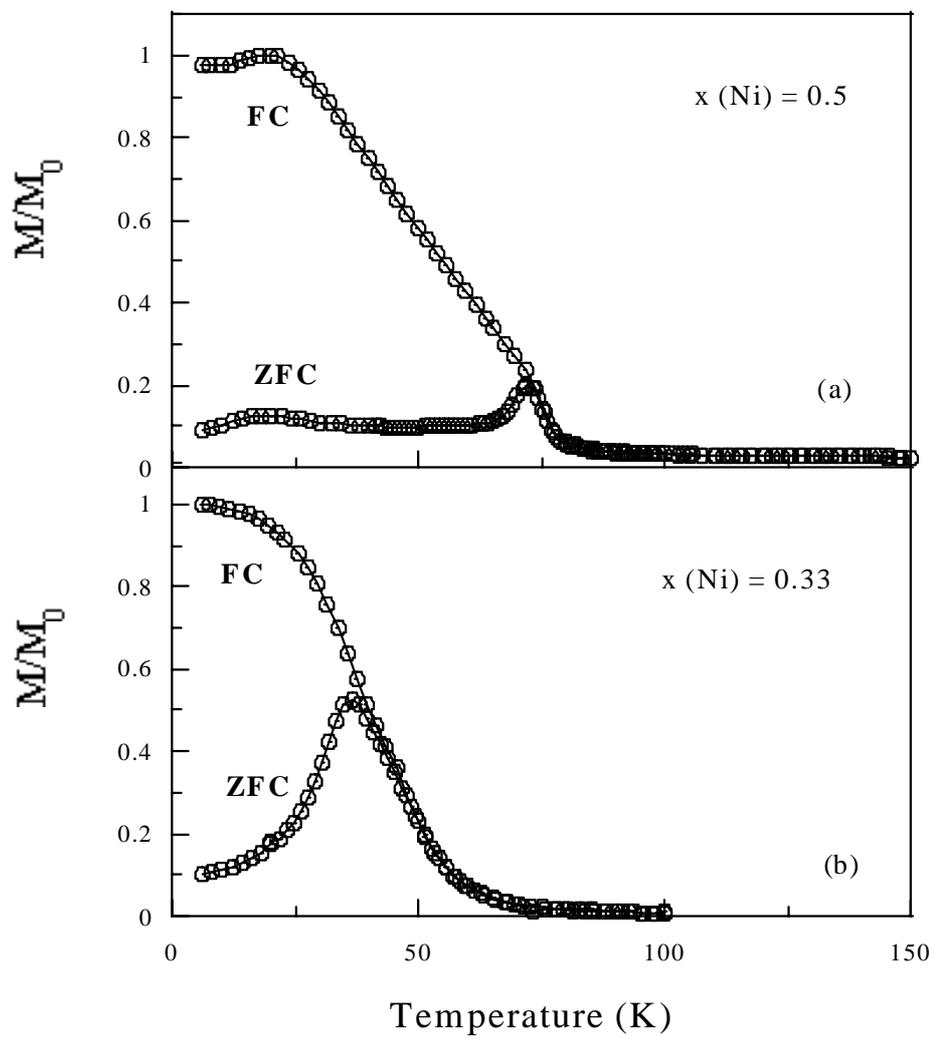

Fig. 3    Ma et al.

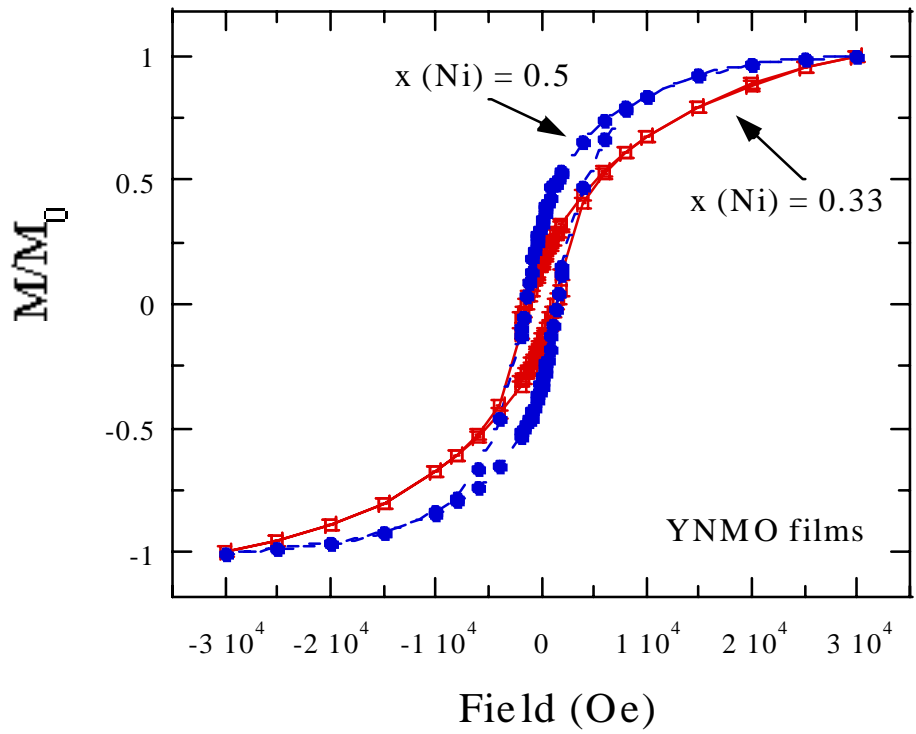

Fig. 4   Ma et al.